# Cycle Inverse-Consistent TransMorph:A Balanced Deep Learning Framework for Brain MRI Registration

Jiaqi Shang, Haojin Wu, Yinyi Lai, Zongyu Li, Chenghao Zhang, Jia Guo

*Abstract*—Deformable image registration plays a fundamental role in medical image analysis by enabling spatial alignment of anatomical structures across subjects. While recent deep learning–based approaches have significantly improved computational efficiency, many existing methods remain limited in capturing long-range anatomical correspondence and maintaining deformation consistency. In this work, we present a cycle inverse-consistent transformer-based framework for deformable brain MRI registration. The model integrates a Swin-UNet architecture with bidirectional consistency constraints, enabling the joint estimation of forward and backward deformation fields. This design allows the framework to capture both local anatomical details and global spatial relationships while improving deformation stability. We conduct a comprehensive evaluation of the proposed framework on a large multi-center dataset consisting of 2,851 T1-weighted brain MRI scans (1507 female) aggregated from 13 public datasets. Experimental results demonstrate that the proposed framework achieves strong and balanced performance across multiple quantitative evaluation metrics while maintaining stable and physically plausible deformation fields. Detailed quantitative comparisons with baseline methods, including ANTs, ICNet, and VoxelMorph, are provided in the appendix. Experimental results demonstrate that CICTM achieves consistently strong performance across multiple evaluation criteria while maintaining stable and physically plausible deformation fields. These properties make the proposed framework suitable for large-scale neuroimaging datasets where both accuracy and deformation stability are critical.

## I. Introduction

DEFORMABLE medical image registration is a fundamental component of neuroimaging analysis, enabling spatial normalization, population-level comparison, and downstream quantitative analyses such as tensor-based morphometry and brain aging studies [1], [2]. By establishing voxel-wise correspondence across subjects, deformable registration provides the foundation for studying structural variability and disease-related anatomical changes in the brain.

Early registration methods were primarily based on classical optimization frameworks, including intensity-based and feature-based approaches[3], [4] that explicitly model similarity measures and regularization terms [5]. These methods were later extended to diffeomorphic formulations, which enforce smoothness, invertibility, and topology preservation of the estimated transformations [6], [7]. Although such approaches can produce physically plausible deformations, the iterative optimization procedure must be performed separately for each image pair, resulting in substantial computational cost and limited scalability for large neuroimaging datasets.

To address these limitations, learning-based deformable registration methods have been introduced [8], [9], in which a parametrized mapping between image pairs is learned from training data and applied efficiently during inference. Convolutional neural networks (CNNs) have been widely adopted in this context and have demonstrated significant improvements in computational efficiency. However, CNN-based architectures primarily rely on local receptive fields, which may limit their ability to capture long-range anatomical correspondence [10]. Consequently, deformation fields produced by CNN-based models may exhibit reduced symmetry or instability in anatomically complex areas.

Inverse-consistent registration has been explored as an effective strategy to improve deformation reliability by enforcing coherence between forward and backward transformations [11]. By constraining the composition of bidirectional mappings to approximate the identity transformation, inverse consistency promotes symmetry and reduces error accumulation in deformable registration. Nevertheless, when inverse-consistency constraints are combined with purely convolutional representations, the estimation of deformation fields may still be limited by insufficient global context modeling.

More recently, transformer-based architectures have been introduced to medical image registration to address the limitation of local receptive fields in convolutional networks [12], [13]. These models leverage hierarchical representations and window-based self-attention mechanisms to capture both local anatomical details and long-range spatial dependencies efficiently. These models have demonstrated improved registration performance and deformation symmetry compared with conventional CNN-based approaches [10], [14]. Building upon these developments, cycle inverse-consistent transformer-based frameworks have emerged as promising solutions for deformable medical image registration by combining global context modeling with bidirectional consistency constraints.

In this work, we present a systematic implementation and comprehensive large-scale evaluation of cycle inverse-consistent TransMorph (CICTM) for deformable brain MRI registration. Our study aims to establish the robustness and scalability of this framework through extensive empirical validation. The main contributions of this work are summarized as follows:

1. Large-scale validation: We conduct extensive evaluation on a multi-center dataset of 2,851 T1-weighted brain MRI scans, providing a large-scale assessment of registration performance across diverse populations.

2. Systematic benchmarking: The proposed framework is compared with representative learning-based methods (VoxelMorph and ICNet) and a classical diffeomorphic approach (ANTs) using multiple complementary evaluation metrics.

3. Robust implementation: We provide a reproducible registration pipeline with detailed implementation specifications, enabling practical deployment for population-level neuroimaging studies.

Overall, this study provides a comprehensive empirical evaluation of the CICTM framework, analyzing registration accuracy, deformation stability, and scalability using multiple similarity metrics [15] and Jacobian-based regularity analysis [16], [17], thereby supporting its application in large-scale neuroimaging studies.

## II. MATERIALS AND METHODS

### A. Dataset and Data Split

A large multi-center dataset consisting of 2,851 T1-weighted structural brain MRI scans was aggregated from 13 publicly available neuroimaging datasets, including ADNI, IXI, MindConnect, AIBL, SALD, BCGSP, CoRR, OASIS, OASIS-3, NIFD, PPMI, SLIM, and OASIS-2. These datasets span multiple acquisition sites and scanners, providing heterogeneous imaging data for evaluating the generalizability of the proposed registration framework. The distribution of samples across datasets is illustrated in Fig. 1. The cohort spans a wide adult age range and includes both male and female subjects, providing a representative population for large-scale neuroimaging analysis.

All images were randomly partitioned into training, validation, and test sets using a fixed split ratio of 8:1:1. To ensure reproducibility, a fixed random seed (seed = 42) was used during dataset splitting. This resulted in 2,280 images for training, 285 images for validation, and 286 images for testing. The same data split was used consistently across all experiments to ensure fair and reproducible comparisons between different registration models.

Overall, the dataset contains 52.9% female and 47.1% male subjects, with similar proportions maintained across the training, validation, and test subsets. The demographic distributions of age and sex for the full dataset and each data split are summarized in Fig. 1, which presents the distributions for the entire dataset as well as for the training, validation, and test subsets.

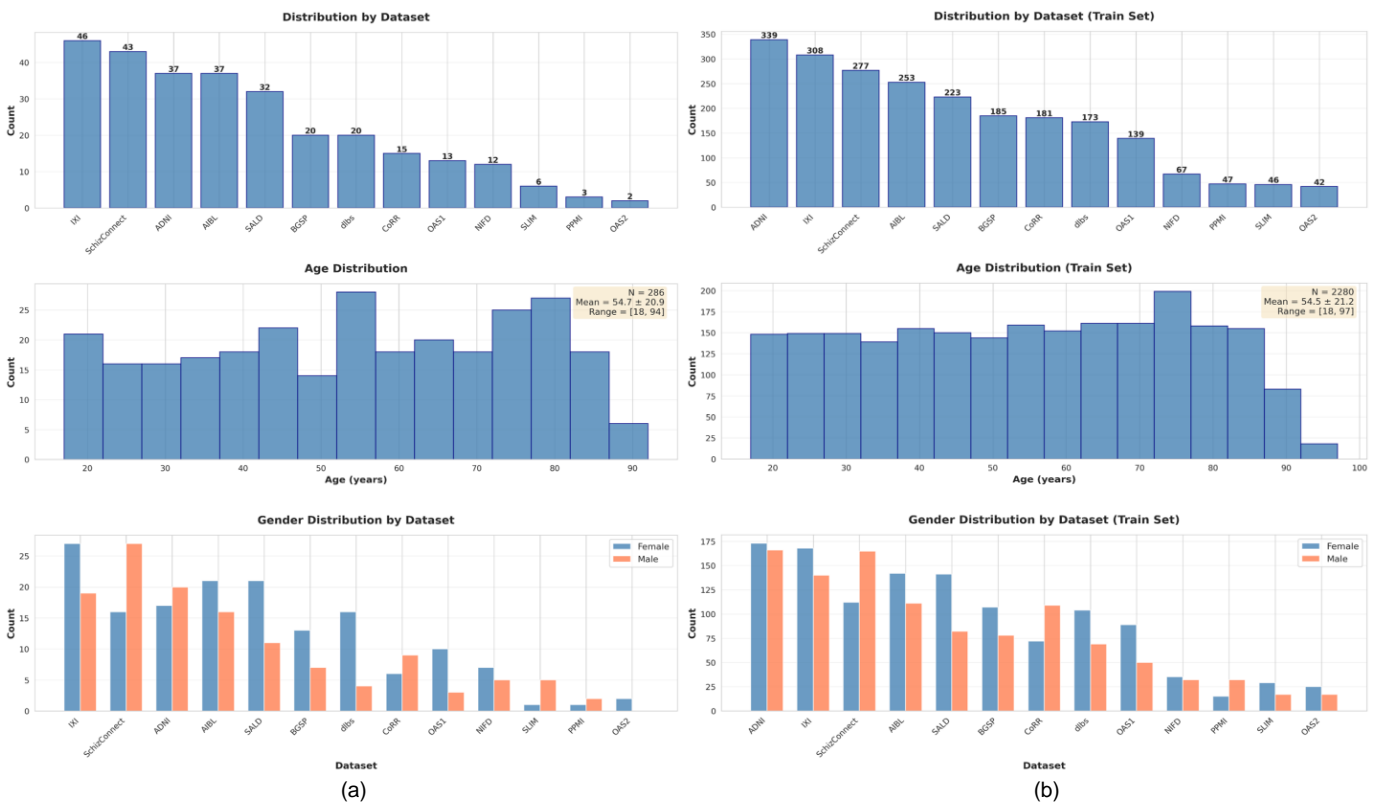

(a)  (b)

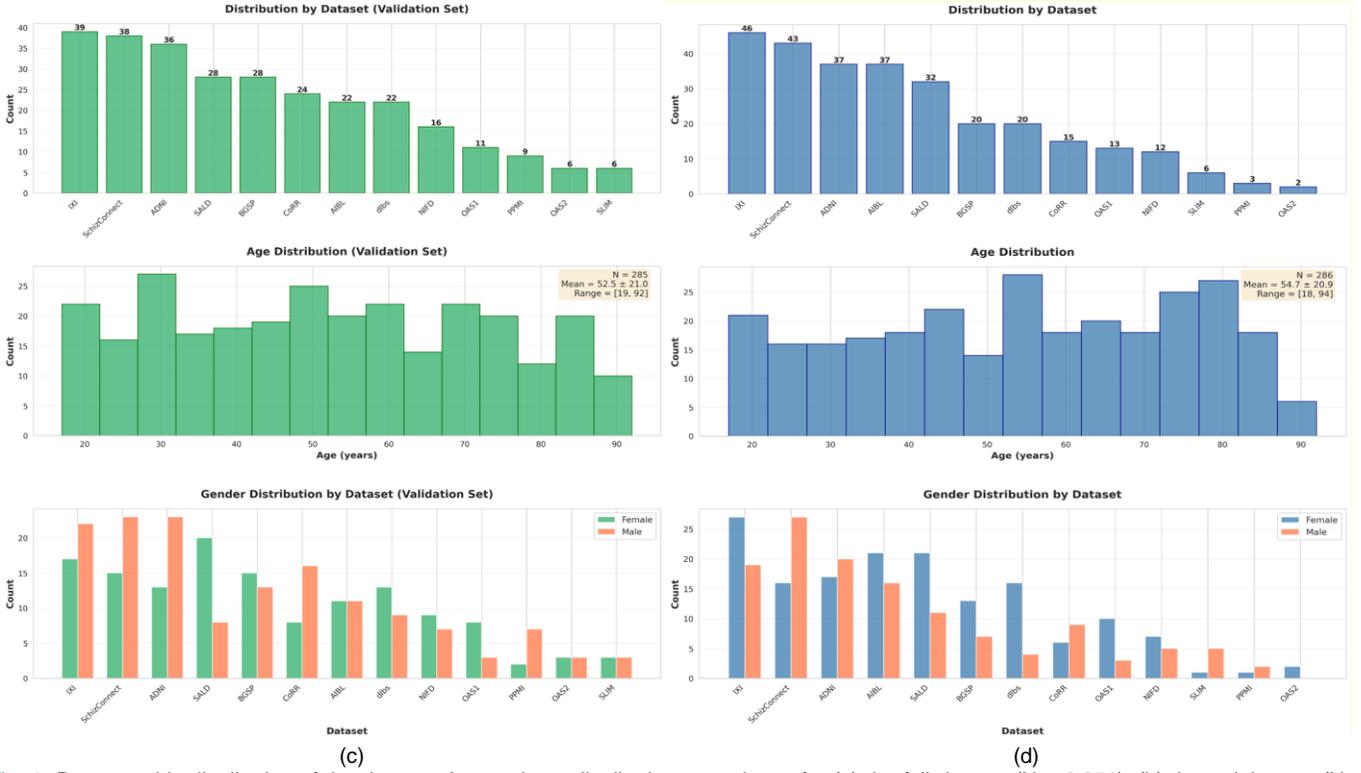

Fig. 1. Demographic distribution of the dataset. Age and sex distributions are shown for (a) the full dataset (N = 2,851), (b) the training set (N = 2,280), (c) the validation set (N = 285), and (d) the test set (N = 286). The dataset was randomly split using an 8:1:1 ratio. Comparable age ranges and sex proportions across subsets indicate that the data split did not introduce noticeable sampling bias.

### B. Preprocessing

All T1-weighted brain MRI scans were preprocessed using a standard neuroimaging pipeline. Bias field inhomogeneity was corrected using the N4 bias field correction algorithm implemented in ANTsPyX (version 0.5.4). To eliminate global pose differences across subjects, all scans were affinely aligned to the MNI152 template prior to deformable registration. No nonlinear spatial normalization was applied during preprocessing to ensure that deformable alignment was learned entirely by the registration models.

All scans were resampled to an isotropic resolution of $1\times1\times1$ mm³ and cropped to a standard field of view of $160\times192\times192$ voxels. For each scan, voxel intensities within the brain mask were min-max normalized to [0, 1]. These preprocessing steps ensure consistent input dimensions and intensity distributions across all subjects.

### C. Network Architecture

The proposed framework adopts a cycle inverse-consistent transformer-based architecture for deformable medical image registration (Fig. 2). The network is built upon a 3D Swin-UNet backbone, where hierarchical Swin Transformer blocks extract multi-scale feature representations from paired input images, followed by a decoder that predicts dense 3D deformation fields.

Given a moving image IM and a fixed image IF, the two images are concatenated and processed by the network to estimate bidirectional deformation fields. A differentiable spatial transformer network (STN) is used to warp the moving image according to the predicted transformation, producing the registered image in the fixed image space.

To support cycle inverse-consistent learning, both forward and backward transformations are estimated during training. This design encourages bidirectional consistency and improves deformation stability.

The network uses an input size of $160 \times 192 \times 192$, a patch size of 4, an embedding dimension of 48, and Swin Transformer depths of [2, 2, 18, 2] for both the encoder and decoder. The numbers of attention heads are [4, 8, 16, 32], and the window size is set to $5 \times 6 \times 6$. A lightweight registration head is attached to the decoder output to generate a three-channel dense deformation field for 3D registration. The architecture is implemented based on the TransMorph framework [12] with additional cycle inverse-consistent learning design.

### D. Inverse-Consistency and Loss Function

To improve registration accuracy and deformation stability, the model was trained using a composite loss function consisting of an image similarity term, a smoothness regularization term, inverse-consistency terms, and a Jacobian determinant penalty.

Image similarity between the warped image and the target image was measured using multi-scale structural similarity (MS-SSIM) loss, which provides a robust measure of structural correspondence across multiple spatial scales. To encourage spatially smooth and anatomically plausible deformations, an L2 smoothness regularization term was applied to the predicted deformation field.

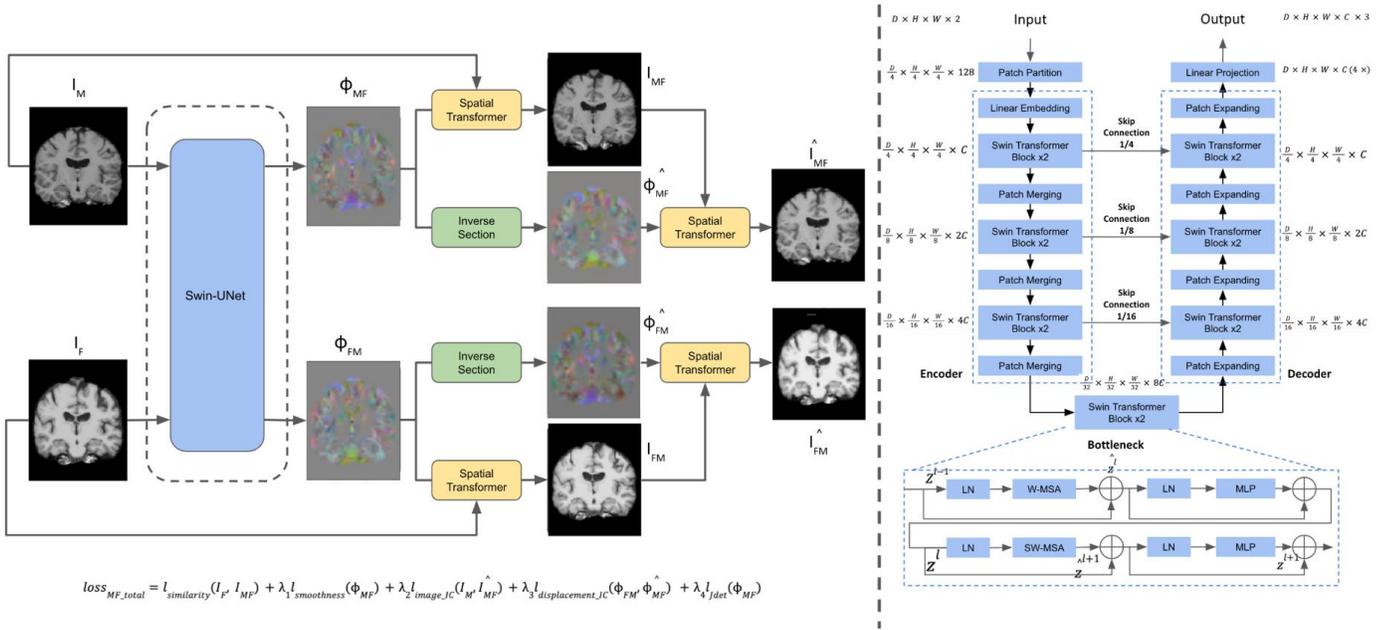

$$loss_{MF\_total} = l_{similarity}(I_F, I_{MF}) + \lambda_1 l_{smoothness}(\phi_{MF}) + \lambda_2 l_{image\_IC}(I_M, \hat{I}_{MF}) + \lambda_3 l_{displacement\_IC}(\phi_{FM}, \hat{\phi}_{MF}) + \lambda_4 l_{Jdet}(\phi_{MF})$$

Fig. 2. Schematic illustration of the cycle inverse-consistent registration framework. The framework jointly estimates forward and backward deformation fields using a Swin-UNet backbone. A spatial transformer network is applied to generate warped images, and inverse modules are used to enforce inverse consistency at both the image and displacement field levels during training.

To support cycle inverse-consistent learning, both forward and backward transformations were estimated during training. Two inverse-consistency terms were used: an image-level inverse-consistency term, which enforces similarity between the reconstructed image and the original image after forward and reverse warping, and a flow-level inverse-consistency term, which promotes consistency between the bidirectional deformation fields.

All loss components were combined using empirically selected weighting coefficients: $\lambda_1$= 0.5 for smoothness regularization, $\lambda_2$= 10 for image-level inverse consistency, $\lambda_3$= 1.0 for flow-level inverse consistency, and $\lambda_4$= 1000 for Jacobian determinant penalty. These coefficients were determined through validation set performance and kept fixed across all experiments. Detailed mathematical formulations are provided in Supplementary Material A.

### E. Inverse-Consistency and Loss Function

At inference time, pairwise deformable registration is performed in a single forward pass of the trained network. Given a moving image $I_M$ and a fixed image $I_F$, the model predicts a dense deformation field $\varphi_{MF}$ (and the reverse field $\varphi_{FM}$). A differentiable STN then warps the moving image using the predicted deformation field to produce the registered image in the fixed image space:

$$I_{MF} = I_M \circ \varphi_{MF}$$

The output of pairwise registration therefore consists of the warped image $I_{MF}$ and its corresponding deformation field $\varphi_{MF}$, which can be used for subsequent analyses such as morphometric measurements or deformation-based statistics.

### F. Baseline Methods

To evaluate the effectiveness of the proposed framework, comparisons are conducted with representative learning-based and classical deformable registration methods, including VoxelMorph, ICNet, and ANTs.

All baseline methods are evaluated using the same dataset splits as the proposed method. Publicly available implementations are employed, and default parameter settings are used unless otherwise specified. No additional tuning is performed to favor any particular method. This ensures a fair and reproducible comparison across all approaches.

### G. Evaluation Metrics

Registration performance is evaluated using a comprehensive set of quantitative metrics that assess image similarity, anatomical alignment accuracy, and local structural consistency.

Image similarity between the warped image and the fixed image is assessed using multiple complementary metrics, including the structural similarity index measure (SSIM), normalized cross-correlation (NCC), mutual information (MI), peak signal-to-noise ratio (PSNR), mean squared error (MSE), and mean absolute error (MAE). These metrics collectively characterize global intensity agreement, local structural correspondence, and pixel-wise reconstruction fidelity.

The Dice similarity coefficient is computed on binarized brain masks derived from intensity thresholding (threshold = 0.1) after intensity normalization. This metric evaluates the spatial overlap between the warped image and the fixed template.

To further assess local structural consistency, gradient similarity is computed between the warped image and the fixed image, providing a measure of edge and boundary alignment that is sensitive to anatomical details.

All metrics are computed on the same test set for all methods using identical evaluation procedures to ensure fair and unbiased comparison.

## III. RESULTS

### A. Overall Registration Performance

Quantitative results are summarized in Table 1 (Quantitative comparison of CICTM with baseline methods, including ANTs, ICNet, and VoxelMorph, evaluated on the test set. Results are reported as mean ± standard deviation across all test image pairs. Higher values indicate better performance for SSIM, NCC, MI, PSNR, Dice and gradient similarity, while lower values indicate better performance for MSE and MAE.). The proposed CICTM consistently achieves the best performance across most evaluation metrics. In particular, it obtains the highest SSIM and NCC values, indicating improved structural similarity and intensity consistency between the warped and fixed images.

Compared with classical ANTs registration and learning-based baselines including VoxelMorph and ICNet, CICTM demonstrates improved alignment accuracy and better preservation of anatomical structures.

In terms of structural similarity, CICTM achieves the highest SSIM score (0.8585 ± 0.0270), indicating superior preservation of anatomical structures compared with all baseline methods. CICTM also attains the highest mutual information and gradient similarity scores, suggesting improved alignment of both global intensity distributions and local structural details.

For anatomical overlap, Dice similarity is computed on binarized brain masks to evaluate global spatial correspondence between the warped image and the template. CICTM achieves a Dice coefficient of 0.9786 ± 0.0059, which is comparable to ANTs and VoxelMorph and substantially higher than ICNet.

VoxelMorph achieves the highest NCC value, while ICNet attains lower MSE and MAE, while achieving higher PSNR. However, improvements in certain intensity-based or error-based metrics do not necessarily correspond to better structural similarity or anatomical alignment, as reflected by the relatively lower SSIM, Dice, and MI scores. Overall, CICTM demonstrates a more balanced performance across complementary evaluation criteria.

TABLE I QUANTITATIVE COMPARISON OF REGISTRATION PERFORMANCE.

| Metric | CICTM | ANTs | ICNet | VoxelMorph |
|---|---|---|---|---|
| SSIM | **0.8585 ± 0.0270** | 0.8222 ± 0.0226 | 0.7692 ± 0.0110 | 0.8416 ± 0.0234 |
| NCC | 0.9752 ± 0.0077 | 0.9677 ± 0.0116 | 0.9456 ± 0.0109 | **0.9760 ± 0.0061** |
| MI | **0.6966 ± 0.0143** | 0.6806 ± 0.0109 | 0.5861 ± 0.0083 | 0.6751 ± 0.0124 |
| PSNR | 17.0728 ± 2.7428 | 16.5527 ± 2.6042 | **18.4837 ± 1.6676** | 17.9466 ± 2.8333 |
| Dice | **0.9786 ± 0.0059** | 0.9781 ± 0.0095 | 0.9556 ± 0.0057 | 0.9776 ± 0.0046 |
| Gradient Similarity | **0.7962 ± 0.0346** | 0.7498 ± 0.0514 | 0.6306 ± 0.0307 | 0.7793 ± 0.0246 |
| MSE | 0.0233 ± 0.0123 | 0.0259 ± 0.0133 | **0.0154 ± 0.0074** | 0.0193 ± 0.0108 |
| MAE | 0.0721 ± 0.0232 | 0.0755 ± 0.0234 | **0.0525 ± 0.0154** | 0.0644 ± 0.0220 |

### B. Deformation Regularity and Topology Preservation

Beyond alignment accuracy, deformation regularity is a critical requirement for deformable medical image registration, particularly in population-level neuroimaging analyses. We evaluate deformation quality by analyzing the Jacobian determinant of the predicted deformation fields.

CICTM produces deformation fields with a substantially lower proportion of non-positive Jacobian determinants compared with learning-based baseline methods, indicating reduced folding and improved topology preservation. This behavior can be attributed to the explicit inverse-consistency constraints and Jacobian determinant regularization that penalizes negative Jacobian determinants incorporated during training.

In contrast, ICNet exhibits a higher incidence of local folding, while VoxelMorph occasionally produces irregular deformations despite achieving strong intensity-based alignment. ANTs generates diffeomorphic transformations by design, resulting in smooth deformation fields, albeit at the cost of increased computational complexity. These results demonstrate that CICTM effectively balances deformation regularity and registration accuracy.

Figure 3 further illustrates the deformation characteristics produced by CICTM. The log-Jacobian determinant maps show smooth spatial transitions without large regions of negative values, indicating effective topology preservation. The corresponding deformation magnitude maps reveal anatomically plausible displacement patterns, with larger deformations localized to cortical and ventricular regions, consistent with expected inter-subject variability. These visual results corroborate the quantitative analysis and demonstrate that CICTM produces physically meaningful deformation fields.

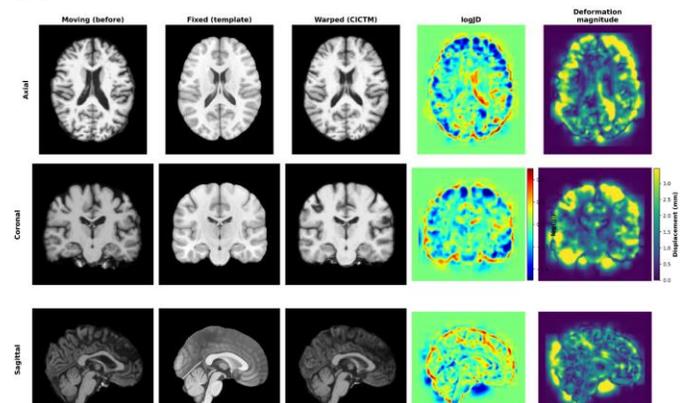

Fig. 3. Deformation analysis of CICTM. Visualization of deformation characteristics produced by CICTM. From left to right: moving image,

fixed template, warped image, log-Jacobian determinant (logJD) map, and deformation magnitude map. Results are shown in axial, coronal, and sagittal views. The logJD maps demonstrate smooth spatial variation without prominent folding, while deformation magnitude maps highlight anatomically plausible displacement patterns.

### C. Qualitative Visualization

Figure 4 presents qualitative comparisons of axial, coronal, and sagittal slices for different registration methods. Compared with ANTs, VoxelMorph, and ICNet, CICTM achieves improved anatomical alignment while preserving sharper structural boundaries. In particular, CICTM demonstrates more consistent cortical alignment and reduced local distortions across all views. These observations are consistent with the quantitative improvements reported in Table 1.

In particular, CICTM better preserves fine anatomical details while avoiding noticeable distortions or folding artifacts. ICNet occasionally exhibits over-smoothing or local misalignment, whereas VoxelMorph may introduce subtle but spatially inconsistent deformations. ANTs produce smooth and plausible deformations but may fail to capture fine-grained anatomical correspondence in certain regions.

These qualitative observations are consistent with the quantitative findings and further highlight the advantages of the proposed method in preserving both anatomical fidelity and deformation plausibility.

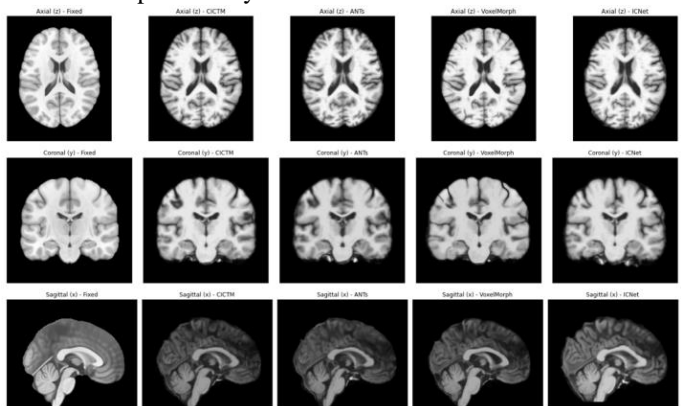

Fig. 4. Qualitative comparison of registration results. Qualitative comparison of registration results across different methods, including CICTM, ANTs, VoxelMorph, and ICNet. Axial, coronal, and sagittal views are shown. CICTM exhibits improved anatomical alignment and sharper structural boundaries compared with baseline methods.

### D. Implications for Population-Level Neuroimaging Analysis

Accurate and topology-preserving deformable registration is essential for population-level neuroimaging analyses, where subtle anatomical variations across subjects are of primary interest. Errors in registration or non-physical deformations may propagate into downstream analyses and lead to biased or unreliable results.

The balanced performance of CICTM across image similarity, anatomical overlap, and deformation regularity metrics suggests that it provides a reliable foundation for large-scale neuroimaging studies. In particular, the improved stability observed across test image pairs, together with reduced folding in the deformation fields, indicates that CICTM can produce consistent and physically plausible spatial normalization.

These properties are especially important for applications that rely on deformation-derived measurements, such as tensor-based morphometry and voxel-wise statistical analysis. By preserving anatomical correspondence while maintaining deformation regularity, CICTM enables robust group-level comparisons and supports reproducible population-level inference.

### E. Computational efficiency

Inference time statistics for all registration methods are summarized in Table 2. (Inference time per image pair for different registration methods evaluated on the test set. Results are reported in seconds and summarized using mean, median, minimum, and maximum values. All methods were evaluated under identical hardware conditions.) All experiments were conducted under identical hardware conditions, using a single NVIDIA RTX 4090 GPU (24 GB) and an Intel Xeon Platinum 8474C CPU with 15 vCPUs, ensuring a fair comparison across methods.

The classical optimization-based method ANTs exhibits substantially longer inference time than all learning-based approaches, with a mean runtime of 55.05 seconds per image pairand a maximum exceeding 60 seconds. This reflects the high computational cost of iterative diffeomorphic optimization.

In contrast, all deep learning–based methods achieve seconds-level inference. VoxelMorph is the fastest method, with a mean inference time of 6.89 seconds, followed by ICNet at 8.38 seconds and the proposed CICTM at 14.68 seconds. Although CICTM is slower than purely convolutional architectures, it remains approximately 3.7× faster than ANTs.

The modest increase in runtime for CICTM can be attributed to the additional computational overhead introduced by Transformer-based global attention and bidirectional cycle-consistency enforcement. Nevertheless, its inference time remains well within practical limits for large-scale population studies, where a balance between computational efficiency and deformation reliability is essential.

TABLE II
INFERENCE TIME COMPARISON OF DIFFERENT REGISTRATION METHODS.

|        | ANTs ($t$) | CICTM ($t$) | VoxelMorph ($t$) | ICNet ($t$) |
|--------|------------|-------------|------------------|-------------|
| Mean   | 55.05      | 14.68       | 6.89             | 8.38        |
| Median | 55.09      | 14.79       | 6.92             | 8.43        |
| Min    | 50.63      | 9.77        | 4.43             | 5.35        |
| Max    | 60.03      | 25.84       | 14.08            | 14.40       |

## IV. DISCUSSION

In this work, we present CICTM, a transformer-based deformable registration framework that explicitly enforces inverse consistency and deformation regularity. Through extensive quantitative and qualitative evaluations, we demonstrate that CICTM achieves a balanced improvement across image similarity, anatomical alignment, deformation plausibility, and computational efficiency.

A key observation from our results is that improvements in pixel-wise similarity metrics do not necessarily translate into improved anatomical correspondence or physically meaningful deformations. While some baseline methods achieve competitive performance on intensity-based or error-based metrics, they may exhibit reduced structural consistency or increased deformation irregularity. By contrast, CICTM consistently performs well across complementary evaluation criteria, suggesting that explicit modeling of inverse consistency and deformation constraints is critical for robust medical image registration.

The integration of a transformer-based architecture enables CICTM to capture long-range spatial dependencies that are difficult to model using purely convolutional networks. This capability is particularly beneficial for brain MRI registration, where anatomical correspondence often depends on global context rather than local appearance alone. At the same time, the use of inverse-consistency constraints mitigates asymmetric deformation behavior and improves the stability of the predicted transformation fields.

Beyond registration accuracy, deformation regularity is a crucial requirement for downstream neuroimaging applications. The reduced incidence of folding and improved Jacobian behavior observed in CICTM indicate that the model produces physically plausible deformations suitable for population-level analyses. This property is especially important for applications such as tensor-based morphometry and voxel-wise statistical analysis, where deformation-derived measures are directly interpreted.

From a computational perspective, CICTM offers a favorable trade-off between efficiency and modeling capacity. Although its inference time is higher than that of purely convolutional architectures, it remains substantially faster than classical optimization-based approaches while providing improved deformation reliability. This balance makes CICTM well suited for large-scale neuroimaging studies, where both computational scalability and deformation quality are essential.

### A. Limitations

Several limitations of this study should be acknowledged. First, the current evaluation focuses on T1-weighted brain MRI, and the generalizability of the proposed framework to other imaging modalities or anatomical regions has not been extensively explored. Second, while inverse consistency and Jacobian-based regularization improve deformation plausibility, they introduce additional computational overhead, which may limit throughput in extremely large-scale settings. Finally, although qualitative and quantitative evaluations indicate improved stability, downstream task-specific validation, such as disease classification or longitudinal change detection, was not explicitly performed.

### B. Future Work

Future work will focus on extending the proposed framework to multimodal and longitudinal registration scenarios, where inverse consistency and deformation stability are expected to play an even more critical role. Further improvements in computational efficiency, such as model compression or attention approximation, may enable faster inference without sacrificing deformation quality. In addition, integrating CICTM into end-to-end neuroimaging analysis pipelines will allow direct evaluation of its impact on downstream statistical and clinical applications.

## V. CONCLUSION

In this study, we proposed CICTM, a transformer-based deformable medical image registration framework that explicitly enforces inverse consistency and deformation regularity. By integrating global contextual modeling with bidirectional cycle-consistency constraints, CICTM achieves balanced improvements in image similarity, anatomical alignment, and deformation plausibility.

Extensive experiments on a large multi-center brain MRI dataset demonstrate that CICTM provides consistently strong performance across complementary evaluation metrics, while maintaining physically meaningful deformation fields and practical computational efficiency. These properties make CICTM well suited for large-scale population-level neuroimaging studies, where robust spatial normalization and deformation reliability are critical. Overall, this work highlights the importance of jointly considering accuracy, symmetry, and physical plausibility in deep learning–based image registration.

## REFERENCES

[1] X. Hua *et al.*, "3D characterization of brain atrophy in Alzheimer's disease and mild cognitive impairment using tensor-based morphometry," *NeuroImage*, vol. 41, no. 1, pp. 19–34, May 2008, doi: 10.1016/j.neuroimage.2008.02.010.

[2] S. Mishra, I. Beheshti, and P. Khanna, "A Review of Neuroimaging-Driven Brain Age Estimation for Identification of Brain Disorders and Health Conditions," *IEEE Rev. Biomed. Eng.*, vol. 16, pp. 371–385, 2023, doi: 10.1109/RBME.2021.3107372.

[3] B. Mishra, U. C. Pati, and U. Sinha, "Modified demons registration for highly deformed medical images," in *2015 Third International Conference on Image Information Processing (ICIIP)*, Dec. 2015, pp. 152–156. doi: 10.1109/ICIIP.2015.7414757.

[4] J. P. W. Pluim, J. B. A. Maintz, and M. A. Viergever, "Mutual-information-based registration of medical images: a survey," *IEEE Trans. Med. Imaging*, vol. 22, no. 8, pp. 986–1004, Aug. 2003, doi: 10.1109/TMI.2003.815867.

[5] S. Yousefi, N. Kehtarnavaz, K. Gopinath, and R. Briggs, "Two-stage registration of substrcutures in magnetic resonance brain images," in *2009 16th IEEE International Conference on Image Processing (ICIP)*, Nov. 2009, pp. 1729–1732. doi: 10.1109/ICIP.2009.5414540.

[6] Avants, Brian B.; Tustison, Nicholas; Johnson, Hans J., "Advanced Normalization Tools (ANTs)," Zenodo, Dec. 2020. doi: 10.5281/ZENODO.5138159.

[7] L. Risser, F.-X. Vialard, R. Wolz, M. Murgasova, D. D. Holm, and D. Rueckert, "Simultaneous Multi-scale Registration Using Large Deformation Diffeomorphic Metric Mapping," *IEEE Trans. Med. Imaging*, vol. 30, no. 10, pp. 1746–1759, Oct. 2011, doi: 10.1109/TMI.2011.2146787.

[8] G. Balakrishnan, A. Zhao, M. R. Sabuncu, J. Guttag, and A. V. Dalca, "VoxelMorph: A Learning Framework for Deformable Medical Image Registration," *IEEE Trans. Med. Imaging*, vol. 38, no. 8, pp. 1788–1800, Aug. 2019, doi: 10.1109/TMI.2019.2897538.

[9] G. Wu, F. Qi, and D. Shen, "Learning-based deformable registration of MR brain images," *IEEE Trans. Med. Imaging*, vol. 25, no. 9, pp. 1145–1157, Sep. 2006, doi: 10.1109/TMI.2006.879320.

[10] T. C. W. Mok and A. C. S. Chung, "Fast Symmetric Diffeomorphic Image Registration with Convolutional Neural Networks," in *2020 IEEE/CVF Conference on Computer Vision and Pattern Recognition*

*(CVPR)*, Jun. 2020, pp. 4643–4652. doi: 10.1109/CVPR42600.2020.00470.

[11] J. Zhang, "Inverse-Consistent Deep Networks for Unsupervised Deformable Image Registration," Sep. 10, 2018, *arXiv*: arXiv:1809.03443. doi: 10.48550/arXiv.1809.03443.

[12] J. Chen, E. C. Frey, Y. He, W. P. Segars, Y. Li, and Y. Du, "TransMorph: Transformer for unsupervised medical image registration," *Med. Image Anal.*, vol. 82, p. 102615, Nov. 2022, doi: 10.1016/j.media.2022.102615.

[13] G. A. Bello *et al.*, "Deep learning cardiac motion analysis for human survival prediction," *Nat. Mach. Intell.*, vol. 1, no. 2, pp. 95–104, Feb. 2019, doi: 10.1038/s42256-019-0019-2.

[14] Y. Zhu and S. Lu, "Swin-VoxelMorph: A Symmetric Unsupervised Learning Model for Deformable Medical Image Registration Using Swin Transformer," in *Medical Image Computing and Computer Assisted Intervention – MICCAI 2022*, L. Wang, Q. Dou, P. T. Fletcher, S. Speidel, and S. Li, Eds., Cham: Springer Nature Switzerland, 2022, pp. 78–87. doi: 10.1007/978-3-031-16446-0_8.

[15] I. Bakurov, M. Buzzelli, R. Schettini, M. Castelli, and L. Vanneschi, "Structural similarity index (SSIM) revisited: A data-driven approach," *Expert Syst. Appl.*, vol. 189, p. 116087, Mar. 2022, doi: 10.1016/j.eswa.2021.116087.

[16] D. Kuang, "On Reducing Negative Jacobian Determinant of the Deformation Predicted by Deep Registration Networks," Jun. 28, 2019, *arXiv*: arXiv:1907.00068. doi: 10.48550/arXiv.1907.00068.

[17] H. Siebert, C. Großbröhmer, L. Hansen, and M. P. Heinrich, "ConvexAdam: Self-Configuring Dual-Optimization-Based 3D Multitask Medical Image Registration," *IEEE Trans. Med. Imaging*, vol. 44, no. 2, pp. 738–748, Feb. 2025, doi: 10.1109/TMI.2024.3462248.